\documentclass[preprint,showpacs,floats,letterpaper,floatfix,groupedaddress,eqsecnum]{revtex4}
\usepackage{amssymb,amsmath}
\usepackage[dvips]{graphicx}

\usepackage{dcolumn,epsfig}

\begin{document}

\title{ Ratchet effect in inhomogeneous inertial systems: I. Adiabatic case}

\author{W. L. Reenbohn and Mangal C. Mahato}
\email{mangal@nehu.ac.in}
\affiliation{Department of Physics, North-Eastern Hill University, Shillong-793022, India.}

\begin{abstract}
Risken's matrix continued fraction method is used to solve the Fokker-Planck equation to 
calculate particle current in an inertial symmetric (sinusoidal) periodic potential 
under the action of a constant force. The particle moves in a medium with friction 
coefficient also varying periodically in space as the potential but with a finite phase 
difference $\phi (\ne n\pi, n=0,1,2, ...)$. The algebraic sum of current with applied 
forces $\pm|F|$ gives the ratchet current in the adiabatic limit. Even though, the effect
of frictional inhomogeneity is weak, the ratchet current shows very rich qualitative
characteristics. The effects of variation of $F$, the temperature $T$, and the average
friction coefficient $\gamma _0$ on the performance characteristics of the ratchet are 
presented.
\end{abstract}

\vspace{0.5cm}
\date{\today}

\maketitle

\section{Introduction}
In equilibrium, it is not possible to get a steady current in a periodic potential  
without the application of external forces. This is in conformity with the second law of 
thermodynamics. However, in recent times it has become a popular theme to obtain such 
currents in nonequilibrium situations. The currents so obtained in asymmetric but 
periodic potentials with no imposed directional preferences in the presence of 
thermal noise are called thermal ratchet currents\cite{feyn, reim}. The usual 
methods one generally employs to realise such currents are: Rock the system periodically 
alternately with constant forces $|F|$ and $-|F|$ for equal short intervals \cite{magn}, 
or flash the potential either periodically or randomly with two unequal (for example, 
zero and nonzero) amplitudes \cite{prost}. However, the method of obtaining current in 
symmetric periodic potentials subject to asymmetric periodic forcings with zero mean per 
period has also been established \cite{mcm1}. In most of the cases the particle motion is 
considered overdamped which is  more suitable to model particle (such 
as a molecular motor) transport in biological systems where large friction plays important 
role \cite{magn, prost, svob}. There is yet another model wherein the friction is 
considered nonuniform in space and in particular periodically varying similar to the 
potential function. The ratchet current could be obtained, in this inhomogeneous case, 
even in symmetric periodic potentials driven periodically and symmetrically in 
time\cite{pareek}. So far such currents are obtained in overdamped\cite{dan1} 
inhomogeneous systems and the possibility of obtaining such currents in underdamped 
systems\cite{mcm2} had only been indicated earlier. This model is very appropriate to 
describe Josephson junctions \cite{falco} (where the frictional inhomogeneity bears an 
exact analogy to the coupling between concomitant Cooper pair
and the quasiparticle tunnelings), and motion of adatoms on 
crystal surfaces\cite{wahn}. In this part (I) of the work we present detailed 
characteristics of the ratchet current in inhomogeneous (nonuniform friction) inertial 
systems when driven adiabatically. And in the next part (II) we shall present results 
for similar systems when driven at finite frequency.

As will be described in the next section, the steady state mean velocity $\bar v$ of a
particle under the action of a constant force $F$ is calculated using the matrix
continued fraction method (MCFM). The algebraic sum 
$\Sigma \bar v$ ($=\bar v(|F|)+\bar v(-|F|)$)
 of these mean velocities $\bar v$ with applied forces $|F|$ and $-|F|$ is termed here 
as the ratchet current in the adiabatic limit. Since the phase difference $\phi (\ne 0,
n\pi)$ between the sinusoidal potential $V(x)$ and the friction coefficient $\gamma(x)$
is the only parameter in the model that breaks the directional symmetry of the problem, 
the ratchet current is expected to be small. However, $\Sigma \bar v$ turns out to be an 
interesting complicated function of $F$, the temperature $T$, and the mean friction 
coefficient $\gamma_0$. This is not surprising and can be understood quite simply on 
physical grounds as discussed elsewhere\cite{wanda}.

In the same work\cite{wanda} the variation of the ratchet current $\Sigma \bar v$ as a 
function of applied force $|F|$ for various values of the average friction coefficient 
$\gamma_0$ at temperature $T=0.4$ has been presented. (The same choice of dimensionless
parameters, $T$, etc, are elaborated in Sec. II below.) It shows that at the smallest
explored $\gamma_0 (=0.035)$, $\Sigma \bar v$ peaks, with value $\simeq -0.08$, close to
$F=0.09 \ll F_c(=1)$, the critical field. The current is thus thermally assisted.  The
variation of $\Sigma \bar v$ is very sharp: Appreciable current could be seen only in
the range $0.05\leq |F|\leq 0.12$. The current strength reduces almost to zero for
$F=0.12$. Beyond this $F$ no meaningful $\bar v$ could be obtained from MCFM. As
$\gamma_0$ is increased the range of $F$, where appreciable $\Sigma \bar v$ results,
widens monotonically. However, the peak current becomes maximum for 
$\gamma_0 \simeq 0.4$, with
maximum current $\simeq -0.28$ at $|F|\simeq 0.75$, and then gradually decreases as
$\gamma_0$ is increased further. By a close observation of the reported figure
(3, ref. \cite{wanda}) one can conclude that $\Sigma \bar v$ at $T=0.4$ will show
nonmonotonic variation with $\gamma_0$ at carefully chosen $F$ values, which as we 
present in the following is, indeed, the case.

For a given applied force $F$, and the temperature $T$ the current $\Sigma \bar v$
initially increases, attains a peak, and then gradually decreases as $\gamma_0$ is 
increased. The result is reminiscent of the well-known result\cite{kramers} obtained 
by Kramers, however, in an entirely different context of particle flux across a 
potential barrier. In all situations $\Sigma \bar v$ increases with $|F|$, attains a 
peak, and gradually decreases. With appropriately chosen parameters $\gamma_0$ and $T$,  
  $\Sigma \bar v$ shows current reversal as well and consequently as $|F|$ 
is increased further $\Sigma \bar v$ changes its direction and increases in magnitude 
eventually again decreases in magnitude after attaining a second peak. Interestingly, 
$\Sigma \bar v$ shows very similar behaviour also as a function of temperature $T$ in 
some parts of the parameter space $(\gamma_0, F)$.

As the value (and direction) of $\Sigma \bar v$ changes one naturally asks about how it  
is correlated with the changes in energies of the system. It turns out that the average 
kinetic energy of the particle increases monotonically with $|F|$, whereas the average
potential energy shows a nonmonotonic behaviour. Again, in this case too we consider
the mean of the average kinetic (as also of the potential) energies for $F=\pm|F|$. The
average kinetic energy overwhelms the potential energy as $F$ is increased forcing 
the average total energy also to increase monotonically. However, the contribution of the 
average potential energy of the particle is reflected in the derivative of the average
total energy with respect to $F$. These derivatives show distinctive behaviour where
$\Sigma \bar v$ peaks as a function of $F$.

In order to be useful for practical purposes the system must have good performance 
characteristics. We calculate the efficiency of energy transduction of this inhomogeneous
inertial system driven adiabatically and symmetrically and moving in a symmetric 
periodic potential. The system yields small current and hence the work done against an 
applied load is also small if at all it is possible to obtain. Naturally, the 
thermodynamic efficiency of the system is also small and it is feasible to obtain only 
in a very small region of parameter space. However, even without an applied load the 
system continuously performs work against the frictional force and in order to take it 
into account a quantity  called Stokes efficiency has been 
defined\cite{wang, seki, parr}. We calculate the Stokes efficiency of our system. 
This efficiency also has a complex behaviour and, in particular, it is nonmonotonic. 
It is also possible, interestingly, to observe increasing Stokes efficiency 
with temperature contrary to what is observed in usual macroscopic heat engines. This 
kind of behaviour has been shown to occur in overdamped inhomogeneous but asymmetrically 
driven systems\cite{raish}.

Admittedly, the quantitative effect of the inhomogeneity is not very large. Nevertheless, 
our investigation  shows that many important results can be obtained, some even 
counterintuitive, albeit at a small but nonnegligible scale, using system inhomogeneity. 
The ratchet current, the efficiency of performance, and so on, can be improved by orders 
of magnitude, if zero mean asymmetric periodic drive is employed instead of a symmetric 
drive. In this case, however, inhomogeneity does not play a decisive role. Similar 
results can be obtained without system inhomogeneity as the asymmetric drive plays the 
major role.

The above mentioned results will be elaborated upon in section III. The paper will be 
concluded with a discussion in section IV.

\section{The matrix continued fraction method for inhomogeneous inertial systems}
The motion of a particle of mass $m$, moving in a periodic potential $V(x)=-V_0 sin(kx)$
in a medium with friction coefficient $\gamma (x)=\gamma_0(1-\lambda sin(kx+\phi))$
($0<\lambda<1$) and subjected to a constant uniform force $F$, is described by the 
Langevin equation\cite{pareek, amj1},
\begin{equation}
m\frac{d^{2}x}{dt^{2}}=-\gamma (x)\frac{dx}{dt}-\frac{\partial{V(x)}}{\partial
x}+F+\sqrt{\gamma(x)T}\xi(t).
\end{equation}
Here $T$ is the temperature in units of the Boltzmann constant $k_B$. The Gaussian 
distributed fluctuating forces $\xi (t)$ satisfy the statistics: $<\xi (t)>=0$,
and $<\xi (t)\xi(t^{'})>=2\delta(t-t^{'})$. For covenience, we write down Eq.(2.1) in 
dimensionless units by setting $m=1$, $V_0=1$, $k=1$ so that, for example, the scaled
$\gamma_0$ is expressed in terms of the old $\sqrt{mV_0}k$, etc. The reduced Langevin 
equation, with reduced variables denoted again by the same symbols, is written as
\begin{equation}
\frac{d^{2}x}{dt^{2}}=-\gamma(x)\frac{dx}{dt}
+cos x +F+\sqrt{\gamma(x) T}\xi(t),
\end{equation}
where $\gamma(x)=\gamma_0(1-\lambda sin(x+\phi))$. Thus the periodicity of the potential
$V(x)$ and also the friction coefficient $\gamma$ is $2\pi$. The barrier height between
any two consecutive wells of $V(x)$ persists for all $F<1$ and it just disappears at 
the critical field value $F_c=1$. The noise variable, with the same symbol $\xi$, 
satisfies exactly similar statistics as earlier. The Fokker-Planck equation corresponding 
to Eq.(2.2) is given as
\begin{equation}
\frac{\partial W(x,v,t)}{\partial t} = {\cal{L}}_{FP}W(x,v,t),
\end{equation}
where the Fokker-Planck operator
\begin{eqnarray}
{\cal{L}}_{FP} & = & -v\frac{\partial}{\partial x}+\gamma_0(1-\lambda sin(x+\phi))
\frac{\partial}
{\partial v}v - (cosx+F)\frac{\partial}{\partial v}+ \nonumber \\
      &    &\gamma_0T(1-\lambda sin (x+\phi))\frac{\partial^{2}}{\partial v^{2}},
\end{eqnarray}
and $W(x,v,t)$ is the probablity distribution of finding the particle at position $x$
with velocity $v=\frac{dx}{dt}$ at time $t$.

It is hard to solve the equation in the general case. However, our aim is to solve Eqs.
 (2.3-4) in the steady state case. We apply MCFM developed by Risken, 
et. al. \cite{risk}, and used by others to suit various circumstances \cite{ferra}. The 
basic idea behind the method is to expand the distribution function $W(x,v,t)$ in series 
in terms of products of (Hermite) functions $\psi_n(v)$, which are explicit functions of 
velocity $v$, and $C_n(x,t)$ (as coefficients) which are explicit functions of position 
$x$ and time $t$:
\begin{equation}
W(x,v,t)=(2\pi T)^{\frac{-1}{4}}e{^\frac{-v^{2}}{4T}}\sum_{n=0}^
{\infty}C_n(x,t)\psi_n(v),
\end{equation}
with
\begin{equation}
\psi_n=\frac{(b^{\dagger})^n\psi_0}{\sqrt{n!}}, 
\end{equation}
and
\begin{equation}
\psi_0=(2\pi)^\frac{-1}{4}T^\frac{-1}{4}e^\frac{-v^2}{4T}.
\end{equation}
It turns out that $\psi_n(v)$ $(n=0, 1, 2, ...)$ are the eigenfunctions of the operator
$-\gamma(x)b^\dagger b$:
\begin{equation}
-\gamma(x)b^\dagger b \psi_n(v) = -\gamma(x) n \psi_n(v),
\end{equation}
where $b$ and $b^\dagger$ are the bosonic-like  annihilation and creation operators 
satisfying
\begin{equation}
bb^\dagger-b^\dagger b=1,
\end{equation}
and defined as
\begin{equation}
b=\sqrt{T}\frac{\partial}{\partial v}+\frac{1}{2}\frac{v}{\sqrt{T}},
\end{equation}
\begin{equation}
b^\dagger=-\sqrt{T}\frac{\partial}{\partial v}+\frac{1}{2}\frac{v}{\sqrt{T}}.
\end{equation}
(Note: $b^\dagger \psi_n(v)=\sqrt{n+1}\psi_{n+1}(v)$ and 
$b \psi_n(v)=\sqrt{n}\psi_{n-1}(v)$.)
  
Substituting (2.5) in the Fokker-Planck equation (2.3) and multiplying from left by 
$e^\frac{v^2}{4T}$ on both sides of Eq. (2.3) and on simplifying, we get,
\begin{equation}
\sum_{m=0}^{\infty}\frac{\partial C_m(x,t)}{\partial t}\psi_m =
\bar{\cal{L}}_{FP}\sum_{m=0}^{\infty}C_m(x,t)\psi_m(v),
\end{equation}
where 
\begin{equation}
\bar{\cal{L}}_{FP}=e^{\frac{v^2}{4T}}{\cal{L}}_{FP}e^{\frac{-v^2}{4T}}.
\end{equation}
The operator $\bar{\cal{L}}_{FP}$ is given by
\begin{equation}
\bar{\cal{L}}_{FP}=-\gamma(x)b^{\dagger} b -bD-b^{\dagger} \hat D,
\end{equation}
where the operators $D$ and $\hat D$ are defined as,
\begin{equation}
D=\sqrt{T}\frac{\partial}{\partial x},
\end{equation}
\begin{equation}
\hat D=\sqrt{T}\frac{\partial}{\partial x} + \frac{V^{'}-F}{\sqrt{T}},
\end{equation}
with the prime on the potential $V$ denoting differentiation with respect to $x$.

Multiplying both sides of Eq.(2.12) from left by $\psi_n$ and integrating over the
velocity variable and using the orthonormality properties of the functions $\psi_n$, we 
get the equation on $C_n(x,t)$:
\begin{equation}
\frac{\partial C_n(x,t)}{\partial t}+
n\gamma(x)C_n(x,t)+\sqrt{n+1}DC_{n+1}(x,t)+\sqrt{n}\hat{D}C_{n-1}(x,t)=0.
\end{equation}
In the stationary (steady) state case, we thus have,
\begin{equation}
n\gamma(x)C_n(x)+\sqrt{n+1}DC_{n+1}(x)+\sqrt{n}\hat{D}C_{n-1}(x)=0
\end{equation}
giving a recursion relation on $C_n(x)$ ($n=0, 1, 2, ...$). The first ($n=0$) of the
recursion relations reveals that $C_1$ is a constant.

In this steady state case, the normalization condition is given by
\begin{equation}
\int_{0}^{2\pi}\int_{-\infty}^{\infty}W(x,v)dx dv = \int_{0}^{2\pi}C_0 dx =1,
\end{equation}
giving a condition on the coefficient $C_0(x)$. The mean velocity $\bar v(F)$, the 
average potential energy $E_{pot}$, and the average kinetic energy $E_{kin}$, are
obtained, respectively, as,
\begin{equation}
\bar v = \int_{0}^{2\pi} \int_{-\infty}^{\infty}v W(x,v)dx dv = 2\pi \sqrt{T}C_1,
\end{equation}
\begin{equation}
E_{pot}=\int_{-\pi}^{\pi}V(x)C_0(x) dx,
\end{equation}
and
\begin{equation}
E_{kin}=<\frac{v^2}{2}>=\frac{1}{2}+\frac{1}{\sqrt{2}}\int_{0}^{2\pi}C_2(x)dx.
\end{equation}
Thus, all the physical quantities of our interest can be calculated provided $C_0$, $C_1$, 
and $C_2$ are known from Eq.(2.18).

Writing $C_n(x) =\sum_{q}C_n^qe^{iqx}$, $V(x)=\sum_{q}V^qe^{iqx}$, and
$\gamma (x)=\sum_{q}\gamma ^q e^{iq(x+\phi)}$, where $i=\sqrt{-1}$ and $q$'s as integer
indices the recurrence relation (2.18) becomes,
\begin{eqnarray}
\lefteqn{0=} \nonumber \\
 & & n\sum_{p}\gamma^{q-p}e^{i(q-p)\phi}C_n^p 
+ \sqrt{n+1}\sqrt{T}\sum_{p}(ip)\delta_{q,p}C_{n+1}^p 
+\sqrt{n}\sqrt{T}\sum_{p}(ip)\delta_{q,p}C_{n-1}^p \nonumber \\ 
& & -\frac{\sqrt{n}}{\sqrt{T}}\sum_{p}F\delta_{q,p}C_{n-1}^p
+\frac{\sqrt{n}}{\sqrt{T}}\sum_{p}i(q-p)V^{q-p}C_{n-1}^p.
\end{eqnarray}

Defining the matrices $\mathbf{\gamma}, \mathbf{D} , \mathbf{1}$, and $\mathbf{V^{'}}$ 
with components
\begin{eqnarray}
(\mathbf{\gamma})^{q,p}=\gamma^{q-p}e^{i(q-p)\phi},\\
(\mathbf{D})^{q,p}=ip\delta_{q,p},\\
(\mathbf{1})^{q,p}=\delta_{q,p},\\
(\mathbf{V^{'}})^{q,p}=i(q-p)V^{q-p},
\end{eqnarray}
and the column matrices $\mathbf{C_n}, n=0,1,2, ..$ with components
\begin{equation}
(\mathbf{C_n})^q=C_n^q,
\end{equation}
the recurrence relation can be written as,
\begin{equation}
n\mathbf{\gamma C_n} + \sqrt{n+1}\sqrt{T}\mathbf{DC_{n+1}} +
\sqrt{n}(\sqrt{T}\mathbf{D}+\frac{1}{\sqrt{T}}[\mathbf{V^{'}}-F\mathbf{1}])\mathbf{
C_{n-1}}=0.
\end{equation}
Introducing the matrices $\mathbf{M_n}, (n=0, 1, 2, ..)$ such that
\begin{equation}
\sqrt{n}\mathbf{DC_n}=\mathbf{M_nC_{n-1}}
\end{equation}
the recursion relation (2.29) for $n=2$ becomes
\begin{equation}
\mathbf{M_2C_1}(=\sqrt{2}\mathbf{DC_2})=
-\mathbf{D}(\mathbf{\gamma}+\frac{\sqrt{T}}{2}\mathbf{M_3})^{-1}
(\sqrt{T}\mathbf{D}+\frac{1}{\sqrt{T}}[\mathbf{V^{'}}-F\mathbf{1}])\mathbf{C_1}.
\end{equation}
Hence, for nonzero $\mathbf{C_1}$, $\mathbf{M_2}$ can be given in terms of $\mathbf{M_3}$
as
\begin{equation}
\mathbf{M_2}=\mathbf{D}(\mathbf{\gamma}+\frac{\sqrt{T}}{2}\mathbf{M_3})^{-1}
(\frac{1}{\sqrt{T}}[F\mathbf{1}-\mathbf{V^{'}} ] - \sqrt{T}\mathbf{D}).
\end{equation}
and similarly, one can write down for $\mathbf{M_3}$, $\mathbf{M_4}$, and so on. And, in 
general,
\begin{equation}
\mathbf{M_n}=\mathbf{D}(\mathbf{\gamma}+\frac{\sqrt{T}}{n}\mathbf{M_{n+1}})^{-1}
(\frac{1}{\sqrt{T}}[F\mathbf{1}-\mathbf{V^{'}} ] - \sqrt{T}\mathbf{D}),
\end{equation}
for $2\leq n\leq N-1$, if the series of recursion relations is terminated at $n=N$ 
($\mathbf{C_n}=\mathbf{0}$ for $n\geq N+1$). Thus, the last ($n=N$) relation becomes,
\begin{equation}
\mathbf{M_N}=\mathbf{D}\mathbf{\gamma}^{-1}(
\frac{1}{\sqrt{T}}[F\mathbf{1}-\mathbf{V^{'}}]-\sqrt{T}\mathbf{D}),
\end{equation}
for $N\geq 2$. Hence, $\mathbf{M_{N-1}}$, $\mathbf{M_{N-2}}$, ... $\mathbf{M_2}$ can be
given entirely in terms of $\mathbf{\gamma}$, $\mathbf{V^{'}}$, and $F\mathbf{1}$.
Using $\sqrt{2}\mathbf{DC_2}=\mathbf{M_2C_1}$, the recursion relation $(n=1)$\\ 
$\mathbf{\gamma C_1}+\sqrt{2}\sqrt{T}\mathbf{DC_2}+
(\sqrt{T}\mathbf{D}+\frac{1}{\sqrt{T}}[\mathbf{V^{'}}-F\mathbf{1}])\mathbf{C_0}=0$ \\
yields
\begin{equation}
\mathbf{C_0}=\mathbf{HC_1},
\end{equation}
where
\begin{equation}
\mathbf{H}=(\frac{1}{\sqrt{T}}[F\mathbf{1}-\mathbf{V^{'}}]-\sqrt{T}\mathbf{D})^{-1}(
\mathbf\gamma + \sqrt{T}\mathbf{M_2}).
\end{equation}
Since $C_1$ is a constant,
\begin{equation}
C_0^q=\sum_{p}(\mathbf{H})^{q,p}(\mathbf{C_1})^p = (\mathbf{H})^{q,0}C_1.
\end{equation}
From the normalization condition (2.19), we obtain $2\pi (\mathbf{C_0})^0 =1$, and hence 
from Eq.(2.36),
\begin{equation}
C_1=\frac{1}{2\pi(\mathbf{H})^{0,0}}.
\end{equation}
Therefore, once $\mathbf{H}$ is calculated $C_1$ is known and hence all the components
$C_0^q=(\mathbf{C_0})^q$ are evaluated.

\section{Numerical results}
In order to calculate the physical quantities of interest, namely, the mean velocity
$\bar v$, the average kinetic and potential energies for given values of $F$, $\gamma_0$,
and temperature $T$ one needs to evaluate the matrix $\mathbf{H}$. For our special case, 
the potential $V(x)$ and $\gamma(x)$ each has only two Fourier components. Therefore, the 
dimension of $\mathbf{H}$ depends on the number $Q$ of Fourier components to be kept for
the coefficients $C_n(x)$. Also, the degree of complication increases with the number
$N$ of recursion relations one requires to be considered.  The acceptable solution will
require the matrix $\mathbf{H}$ to remain essentially unchanged as $Q$ and $N$ are 
changed. All these numbers depend on the values of $F$, $\gamma_0$, and the temperature 
$T$. Typically, in the worst case, it is sufficient if $Q<30$ and $N<2000$ is taken for 
small $\gamma_0 \ll 1$ in the appropriate parameter space range of ($F, T$). Outside this 
range no 
acceptable solution is found. It turned out that the paramater space $(\gamma_0, F, T)$ 
range where solutions could be extracted was just the region that yielded finite ratchet 
current $\Sigma \bar v$. In all our calculations for which the results are described in 
the following we have set $\phi=0.35$ and $\lambda=0.9$.

\subsection{The ratchet current}
\begin{figure}[]
\newpage
\epsfxsize=9cm
\centerline{\epsfbox{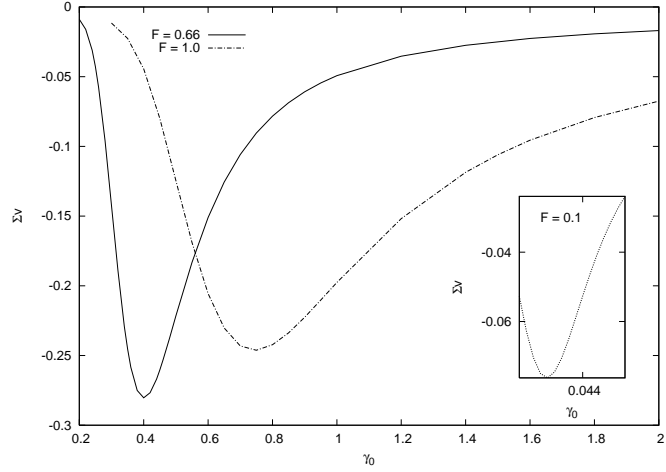}}
\caption{Shows the ratchet current $\Sigma{v}$ as a function of $\gamma_{0}$ for two 
values of $F$. The inset shows the variation of current vs $\gamma_{0}$ for $F =$ 0.1.}
\label{fig.1}
\end{figure}
\begin{figure}[]
\newpage
\epsfxsize=9cm
\centerline{\epsfbox{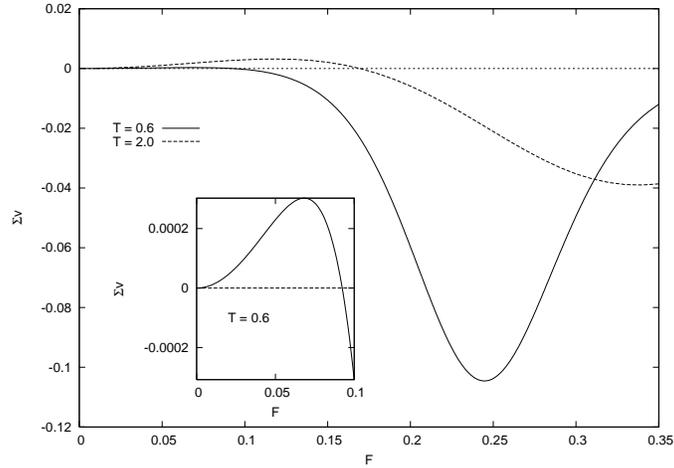}}
\caption{Ratchet current versus F is shown for $\gamma_0=0.1$ and two values of temperature. 
The inset shows variation of $\Sigma{\bar v}$ for smaller values of $F$ at $ T = $0.6. The 
dotted line for zero current is shown just to guide the eye.}
\label{fig.2}
\end{figure}
\begin{figure}[]
\newpage
\epsfxsize=9cm
\centerline{\epsfbox{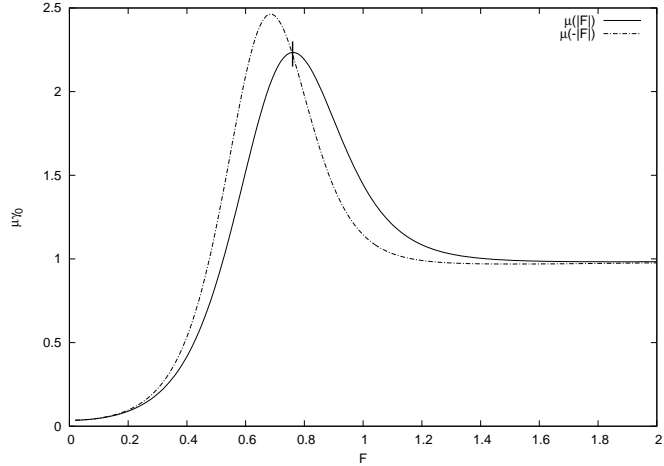}}
\caption{Shows the variation of $\mu(|F|)$ and $\mu(-|F|)$ with $F$ for $\gamma_0 =$ 0.5. 
The little vertical marking ($|$) indicates the position of current maximum.}
\label{fig.3}
\end{figure}
\begin{figure}[]
\newpage
\epsfxsize=9cm
\centerline{\epsfbox{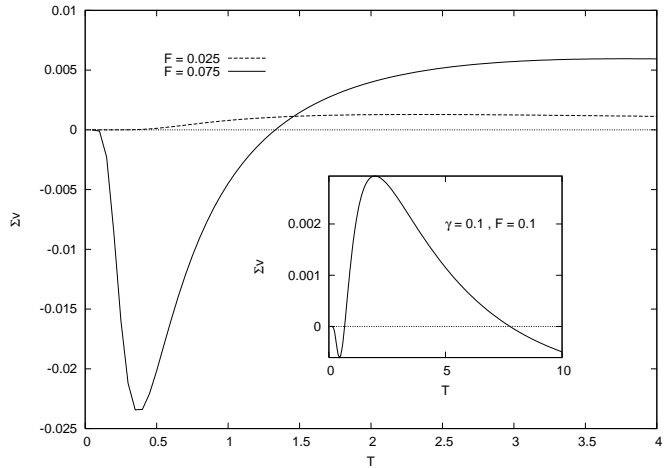}}
\caption{The variation of ratchet current is shown for $\gamma_0 =$ 0.035 for different 
values of $F$. $\Sigma \bar v$ decreases slowly with $T(>4)$. The inset shows two current 
reversals for $\gamma_0 =$ 0.1, $F= 0.1$. The zero $\Sigma \bar v$ line is shown to 
guide the eye.}
\label{fig.4}
\end{figure}

Figure 1 shows how $\Sigma \bar v$ changes with $\gamma_0$ at $T=0.4$. Contrary to usual 
perception current increases as the system becomes more damped. This trend is seen for 
lower $\gamma_0$ values. However, in the higher range of $\gamma_0$, as expected, the 
current diminishes with damping. $\bar v(|F|)$ and $\bar v(-|F|)$ are both monotonically 
decreasing in magnitude with $\gamma_0$. It is only their algebraic sum $\Sigma \bar v$
that shows the nonmonotonic behaviour. It is purely an effect brought about by the 
symmetry breaking due to nonzero phase difference $\phi$ and any analogy with potential 
barrier crossing phenomenon at low damping described by Kramers is superfluous.

In Fig.2, the variation of $\Sigma \bar v$ as a function of $|F|$ for $\gamma_0=0.1$ is
shown for two values of temperature $T=0.6$ and $2.0$. It is to be noted that $T=2.0$
corresponds to an energy exactly equal to the potential barrier height at $F=0$. One 
can see that in the scale of the figure the current $\Sigma \bar v$ remains negative for 
$T=0.6$, whereas at $T=2.0$ the current is positive at low $|F|$ but it changes sign and 
remains negative at higher $|F|$ values for the same phase lag $\phi$. This current 
reversal, however, is neither confined to higher $T$ nor to
higher $\gamma_0$ values. The inset of the figure shows that at smaller $|F|$ region 
$\Sigma \bar v$ is positive and then it becomes negative as $|F|$ is increased. Of 
course, in this case the positive current is an order of magnitude smaller and hence is 
not visible in the larger figure. A similar result is obtained for $\gamma_0=0.035$ where
again the negative current at larger $|F|$ dominate but positve currents at low $|F|$
region are also appreciable. Here again $\bar v(|F|)$ and $\bar v(-|F|)$ both increase
monotonically with $|F|$ only their relative values change giving rise to nonmonotonic 
behaviour of $\Sigma \bar v$, including its direction reversal. At 
higher values of $|F|$ the current should saturate with a small value since the mobility 
$\mu = |\frac{dv}{dF}|$ for both $\pm F$ tend to $\gamma_0^{-1}$. The variation of $\mu$ 
as a function of $|F|$ is shown in Fig.3. It shows that at the position of the current 
maximum $\mu(|F|)=\mu(-|F|)$ as it should be.

The variation of $\Sigma \bar v$ as a function of temperature $T$ at various values of
applied force $|F|$, as shown in Fig.4, reveals that current can also increase with 
temperature. Moreover, one can achieve current reversals as well. In Fig.4, plots are given 
for $\gamma_0=0.035$ and $|F|=0.025$, and 0.075. $|F|=0.075$ yields large currents and shows one current reversal. For larger $|F|$ the current remains negative for all $T$. This 
is not shown here to preserve clarity of the figure. However, in order to highlight the 
existence of two current reversals we plot the graph for $\gamma_0=0.1$, and $|F|=0.1$ 
in the inset of the figure. For low values of $|F|$, for example $|F|=0.025$, the 
current always remains positive, though small, which is consistent with the observation 
made with reference to Fig.2. It is noteworthy that the magnitude of current obtained 
here is quite large compared to what has been reported\cite{raish} for overdamped 
systems.

\subsection{The energetics}
\begin{figure}[]
\newpage
\epsfxsize=9cm
\centerline{\epsfbox{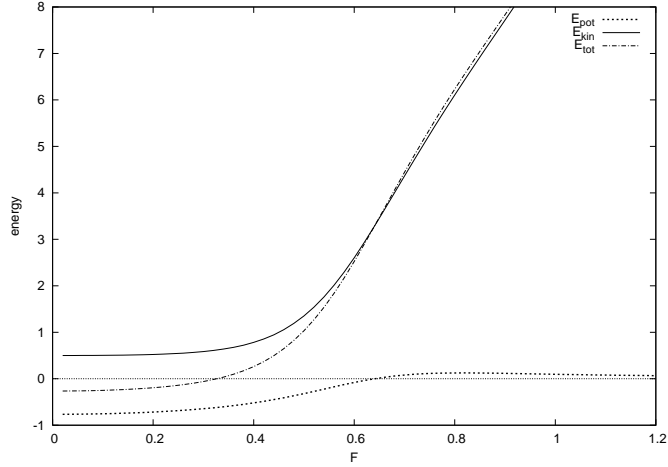}}
\caption{Shows the variation of the average kinetic, potential and total energies at 
$T= 0.4$ for $\gamma_0 = 0.4$ as a function of $F$.}
\label{fig.5}
\end{figure}
\begin{figure}[]
\newpage
\epsfxsize=15cm
\centerline{\epsfbox{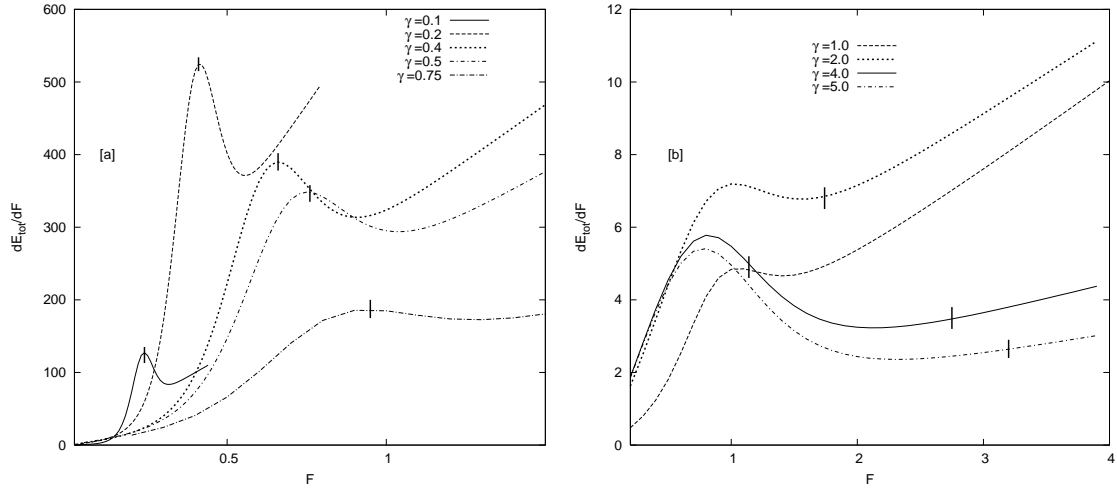}}
\caption{In fig.$[a]$ the slopes of the average total energies are shown for various 
values of $\gamma_0 < 1.0$. The graphs for $\gamma_0 =$ 0.2, 0.4, 0.5 and 0.75 
are scaled by factors of 10, 20, 25 and 30, respectively, to make them visible in the
figure with the graph for $\gamma_0 =$0.1. Fig.$[b]$ shows plots for $\gamma_0 > 1.0$. 
The slopes for $\gamma_0 =$ 2.0, 4.0 and 5.0 are blown up by 4, 5 and 5 times, 
respectively, to scale with the curve for $\gamma_0 =$1.0. The little vertical markings 
on the curves show positions of the current maxima.}
\label{fig.6}
\end{figure}
\begin{figure}[]
\newpage
\epsfxsize=9cm
\centerline{\epsfbox{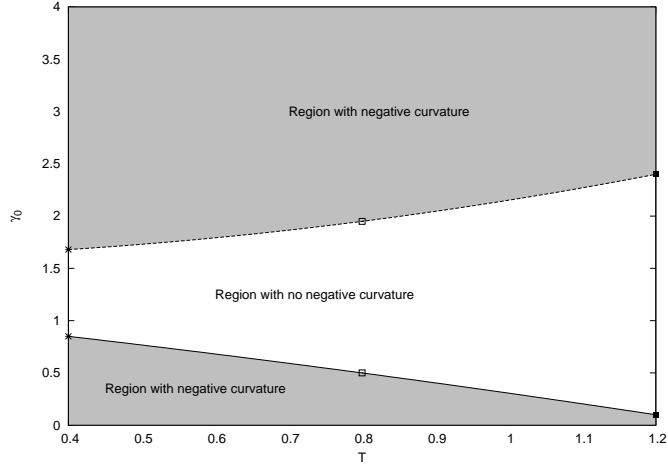}}
\caption{$T-\gamma_{0}$ plane showing two disjointed regions with possibility of negative curvatures of the total energy $E_{tot} (|F|)$ separated by a region without the negative curvature of  $E_{tot} (|F|)$.}
\label{fig.7}
\end{figure}
\begin{figure}[]
\newpage
\epsfxsize=9cm
\centerline{\epsfbox{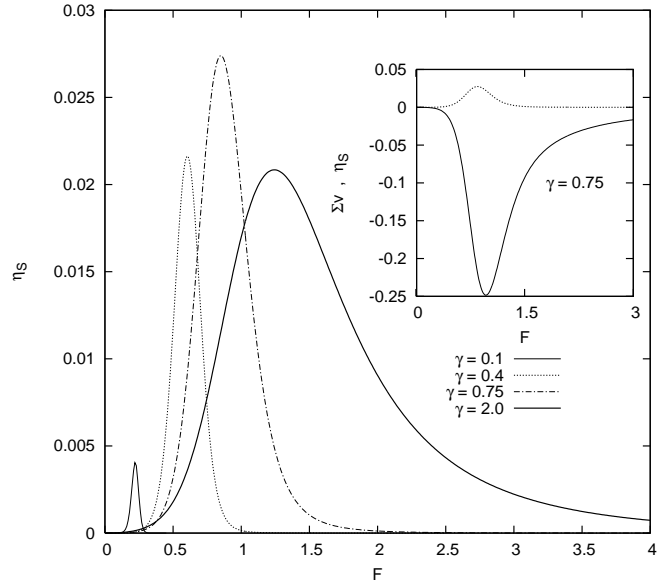}}
\caption{Stokes efficiency versus F for different values of $\gamma_0$. The inset shows 
$\eta_{Stokes}$ and $\Sigma{v}$ plotted for $\gamma_0 =$ 0.75.}
\label{fig.8}
\end{figure}
\begin{figure}[]
\newpage
\epsfxsize=9cm
\centerline{\epsfbox{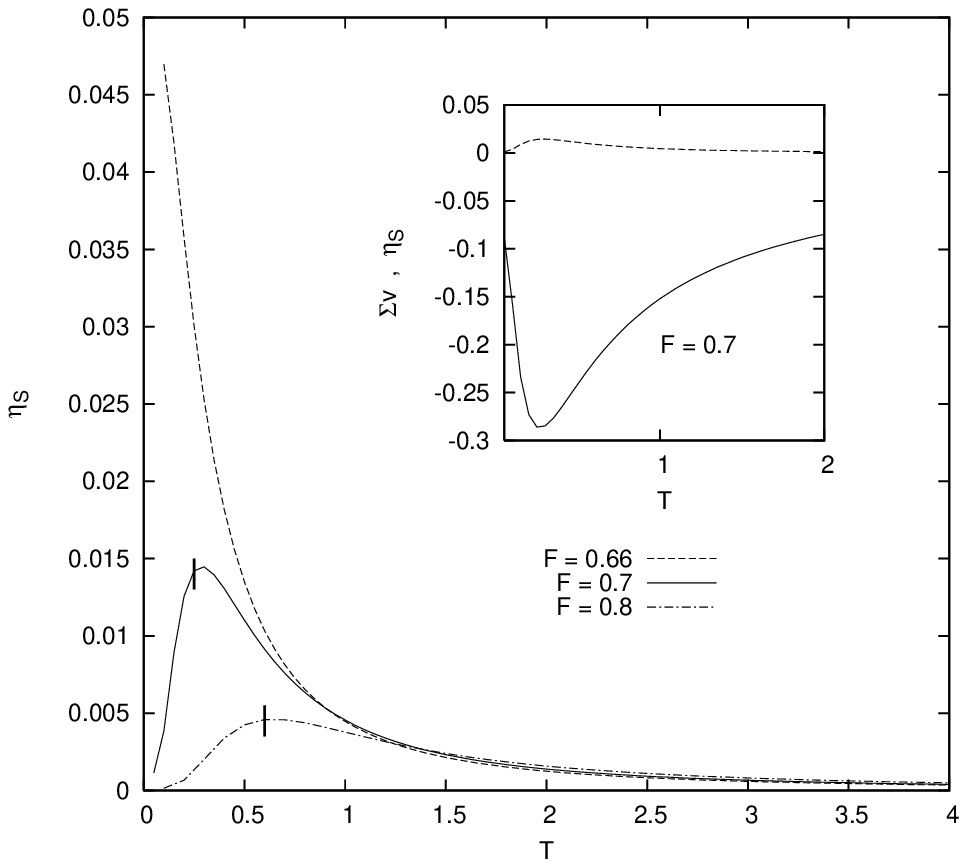}}
\caption{Shows the variation of Stokes efficiency with $T$ for various values of  $F$ 
for $\gamma_0 =$ 0.4. The little vertical markings on the efficiency curves indicate the 
positions of the current maxima. The inset shows efficiency and current for $F =$ 0.7}.
\label{fig.9}
\end{figure}

The average potential and kinetic energies can be easily calculated. The expressions
are given in the Appendix. Fig.5 shows the variation of the average kinetic ($E_{kin}$),
potential ($E_{pot}$), and total ($E_{tot}$) energies at $T=0.4$ for $\gamma_0=0.4$ as a
function of $|F|$. Monotonically increasing nature of $E_{kin}$, and $E_{tot}$ and
nonmonotonic behaviour of $E_{pot}$ are seen for all (explored) $\gamma_0$ and $T$ values.
Interestingly, the slope of $E_{tot}$ shows a shoulder peak which after reaching a minimum
rises again, Fig. 6. For a given temperature the shoulder peak occurs for small $\gamma_0$ 
and large $\gamma_0$ values. However, there is an intermediate range of $\gamma_0$ 
values where the slope does not show this peaking behaviour; the total energy does not
acquire negative curvature in this range of $\gamma_0$. The slope gives valuable 
information about the variation of the ratchet current $\Sigma \bar v$ as shown in Fig.6.
For small $\gamma_0 < 1$ values the slope peaks at an intermediate value of $|F|$, and 
this is precisely where the current maximum occurs. For $\gamma_0 > 1$ the positions of 
the current maximum and that of the slope of the total energy are clearly well separated.
In fact, the current maximum occurs, for these large $\gamma_0$ values, even beyond the 
position of the minimum (after the peak) of the slope of the total energy. At this point 
it is worth mentioning that for not too large temperatures for $\gamma_0 < 1$ the current 
peak occurs at $|F|<1$ whereas for all $\gamma_0 > 1$ the current peaks invariably occur 
at $|F|>1$. Recall that $|F|=1$ is the critical field at which the potential barrier to 
motion just disappears.
A careful observation of the results at various temperatures $T$, thus, allows one to
construct a diagram in the ($T-\gamma_0$) plane, Fig.7, with two disjointed regions with 
possibility of negative curvatures of $E_{tot}(|F_0|)$ separated by a region with only 
positive curvatures. In the "negative curvature" region with $\gamma_0<1$ the slope of 
$E_{pot}$ peaks close to where $\Sigma \bar v$ becomes maximum. Whereas in the other 
negative curvature region with $\gamma_0>1$ current maximum occurs far away from the 
position ($F_0$) where the slope of $E_{tot}(|F_0|)$ peaks. It shows that one can 
clearly separate low damping ($\gamma_0<1$) region from the high damping ($\gamma_0>1$) 
case with each showing qualitatively different characteristic behaviour of particle motion. 

\subsection{The efficiency of performance}
The currents obtained by application of symmetric drives $\pm|F|$ are small and not 
sustainable against applied loads of perceptible magnitude. Therefore it is not feasible 
to calculate work done by the system against the load and hence it is difficult to study 
its behaviour as the other parameters vary. However, the particle keeps moving, essentially 
with average velocity $\Sigma \bar v$, against the frictional force and hence performs 
some work, though not useful practically. How efficiently the system does this work is 
obtained by comparing the magnitude of this work ($=\gamma_0(\sum \bar v)^2$) with the 
input energy supplied to it. A measure of this quantity called the Stokes efficiency is 
defined and given (in our case) as\cite{wang, seki, parr, raish},
\begin{equation}
\eta_{S}=\frac{\gamma_0(\sum \bar v)^2}{|F||(\bar v(|F|)-\bar v(-|F|))|}.
\end{equation}    
In Fig.8, we plot $\eta_{S}$ as a function of $|F|$ for various $\gamma_0$ values at 
$T=0.4$. $\eta_{S}$ behaves in a very similar fashion as the corresponding current does. 
However, $\eta_{S}$ and current do not peak at the same value of F as will be clear from 
the inset of the figure where $\eta_{S}$ and $\Sigma \bar v$ are plotted together for 
$\gamma_0=0.75$. It is remarkable that with such a modest approach one can find regions 
in the parameter space $(\gamma_0,|F|)$ where the efficiency of performance exceeds two 
per cent. It shows that exploiting system inhomogeneity to obtain ratchet current in 
inertial systems may be useful for practical purposes too.

The variation of Stokes efficiency $\eta_{S}$ with $T$ is shown in Fig.9 for $|F|=
0.66, 0.70$, and 0.80 for which  the efficiency and current are appreciable. It is 
noteworthy that at low temperatures the performance of the system improves with rise in 
temperature and only after attaining a peak it declines with further rise in temperature. 
In this case, however, the efficiency and current seem to peak at the same temperature as 
can be seen from the inset of the figure. For $F=0.66$ we could not see the low temperature
behaviour because we do not get sensible results in this region from MCFM.

\subsection{The asymmetrically driven ratchet}
\begin{figure}[]
\epsfxsize=9cm
\centerline{\epsfbox{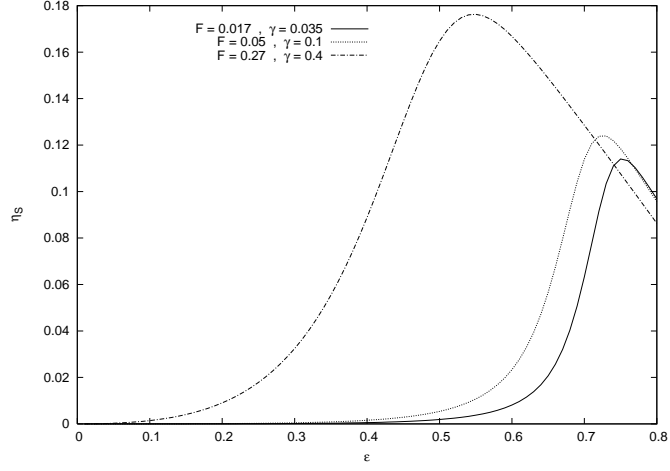}}
\caption{Stokes efficiency as a function of the asymmetry parameter $\epsilon$ for 
different combinations of values of $F$ and $\gamma_0$.}
\label{fig.10}
\end{figure}
\begin{figure}[]
\epsfxsize=9cm
\centerline{\epsfbox{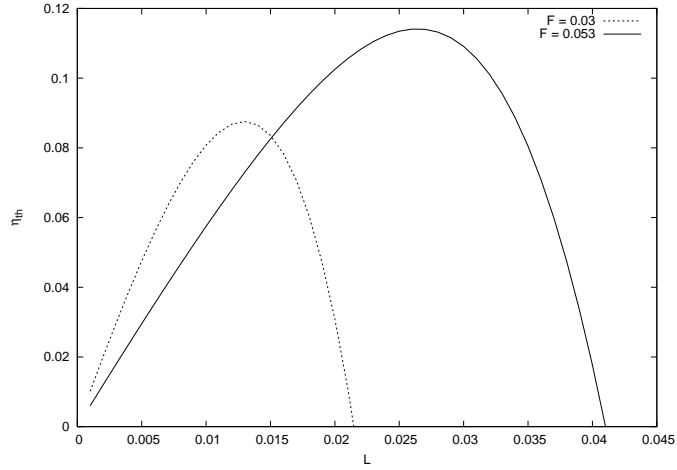}}
\caption{The thermodynamic efficiency  $\eta_{th}$  versus load for two values of $F$ with 
fixed  $\epsilon =$ 0.5, $\gamma_0 =$ 0.035 and $T =$ 0.4.}
\label{fig.11}
\end{figure}

It is clear that in general the efficiency of performance of the inhomogeneous 
inertial ratchets is small except in some special sections of the parameter space
$(\gamma_0, |F|, T)$. This efficiency can be dramatically enhanced if the ratchet is 
driven asymmetrically with an asymmetry parameter $\epsilon (0\leq \epsilon<1)$. Here the 
driving force in the positive and in the negative directions are not same: The force
$F=\frac{1+\epsilon}{1-\epsilon}|F|$ is applied in the positive direction for a fraction
of time $\tau=\frac{1-\epsilon}{2}$ whereas $F=-|F|$ is applied for a fraction of time
$1-\tau=\frac{1+\epsilon}{2}$. The ratchet current now will be given by
\begin{equation}
\Sigma \bar v=\bar v^{+}+\bar v^{-}
\end{equation}
with
\begin{equation}
\bar v^{+}=\frac{1-\epsilon}{2}\bar v(\frac{1+\epsilon}{1-\epsilon}|F|),
\end{equation}
\begin{equation}
\bar v^{-}= \frac{1+\epsilon}{2}\bar v(-|F|)
\end{equation}
The Stokes efficiency of the thermal ratchet motion is similarly defined as in Eq. (3.1):
\begin{equation}
\eta_{S}=\frac{\gamma_0 (\sum \bar v)^2}{
|F||\bar v^{+}(\frac{1+\epsilon}{1-\epsilon})-\bar v^{-}|}.
\end{equation}
In this case the ratchet current, in a typical situation, is quite large. 
Correspondingly, the Stokes efficiency also become large. Fig.10 shows the variation of 
$\eta_{S}$ as a function of the asymmetry parameter $\epsilon$ for various values of
 (a) $|F|=$0.017, $\gamma_{0}=$0.035; (b) $|F|=$0.05, $\gamma_{0}=$0.1 and (c) $|F|=$0.27, $\gamma_{0}=$0.4  at  $T=0.4$. Clearly, more than 10 per cent of ratchet operation 
efficiency becomes achievable. Since the currents are large 
one can also drive these ratchet currents against applied loads, $L$. Hence, the 
thermodynamic efficiency $\eta_{th}$, defined in this case as \cite{seki, parr, raish},
\begin{equation}
\eta_{th}=\frac{L\sum \bar v}{|F||\bar v^{+}(\frac{1+\epsilon}
{1-\epsilon})-\bar v^{-}|} 
\end{equation}
can be calculated. The variation of $\eta_{th}$ as a function of $L$ for $\gamma_0=0.035$
at $T=0.4$ is shown in Fig.11 for $|F|=0.03$, and 0.053. From the figure one can see
that thermodynamic efficiency can be substantial. One can find an optimum load, $L$, and 
an optimum asymmetry parameter, $\epsilon$, of field drive where the ratchet performs most
efficiently. It should, however, be noted that the effect of asymmetric drive is 
overwhelming and the system inhomogeneity plays only a secondary role in this case. One 
can achieve almost similar results without invoking system inhomogeneity as hs been seen 
in the overdamped case\cite{raish}. 
  
\section{Discussion and Conclusion}

All the results of inertial particle motion, in a symmetric periodic potential and driven
symmetrically ( and of course adiabatically), such as ratchet current and its 
nonmonotonic behaviour (including its direction reversals) as a function of $|F|, T$, 
and $\gamma_0$ are possible here only because the friction was considered space-dependent. 
The periodic variation of the friction in space is just a model and we have not provided 
any microscopic basis for it. However, it has been shown earlier that in order to describe
the motion of an adsorbed particle on a crystalline surface (of same atoms as the 
adsorbed one) by a Fokker-Planck equation suitable for one-dimensional particle motion,
one necessarily have to consider space-dependent friction\cite{wahn}. It turned out from
their microscopic theory that the space dependence has to be periodic as the average
potential created by the crystalline (surface) substrate for the adatom and in antiphase
($\phi=\pi$) with it. In other words, the friction will  be large where the substrate 
potential is small. The possiblity of obtaining current due to diffusion 
inhomogeneity\cite{landa} and frictional inhomogeneity\cite{amj1} has been argued 
earlier. The phase difference $\phi\ne n\pi (n=0, 1, 2, ...)$ is necessary to break the 
spatial symmetry in order to get ratchet current. We have found that the ratchet current 
depends on the value of $\phi$. For instance, taking $\phi=0.4$ gives different and 
larger current from what we have reported here for $\phi=0.35$ at the same point in the 
parameter space of ($|F|, T, \gamma_0$). 

The increase of ratchet current and also the efficiency of the ratchet as a function of
$T$ at small temperature range is because the particle has a large potential to surmount 
at small $|F|$ and increasing temperature helps in achieving that. However at the larger
temperature range obtaining a second current peak as shown in Fig.4 cannot be explained
as simply and it arises because of the competition between the temperature and friction
effect as argued elsewhere\cite{wanda}.

The behaviour of the slope of the total energy vis-a-vis the position of ratchet current
peak shows a clear demarcation occurring around  $\gamma_0=1$, in the ($T-\gamma_0$) 
plane. The investigation of time scales of running and locked states of the particle 
motion may provide a clue to this remarkable behaviour. We are carrying out numerical 
Langevin dynamic simulation on this aspect of the problem and the results will be 
reported elsewhere. We conclude by reiterating that particle motion in inhomogeneous 
systems has the potential of a rich field of research.  
\section*{ACKNOWLEDGEMENT}

The authors thank BRNS, Department of Atomic Energy, Govt. of India, for partial
financial support. MCM thanks A.M. Jayannavar for discussion, and Abdus Salam ICTP,
Trieste, Italy for providing opportunity to visit where the paper was written.

\section*{APPENDIX}
Using Eqs. (2.36) and (2.37), the potential energy expression (2.21) is given by
\begin{displaymath}
E_{pot}=\sum_{q}V^{(-q)}\frac{(\mathbf H)^{q,0}}{(\mathbf H)^{0,0}}.
\end{displaymath}
The kinetic energy expression (2.22) can be written as
\begin{equation}
E_{kin}=\frac{1}{2}+{\sqrt{2}}\pi(\mathbf {C_2})^0. \nonumber
\end{equation}
From Eqs. (2.29) and (2.32) one can write down a general equation relating 
$\mathbf{C_n}$ and $\mathbf{C_{n-1}} (n\geq 1)$ as
\begin{displaymath}
\mathbf{C_n}=\frac{1}{\sqrt{n}}\mathbf{A_nC_{n-1}},
\end{displaymath}
where the matrices $\mathbf{A_m} (m=0, 1, 2, ...)$ are given by
\begin{displaymath}
\mathbf{A_m}=(\gamma+\sqrt{T}(m+1)^{-1}\mathbf{DA_{m+1}})^{-1}(
\frac{1}{\sqrt{T}}[F\mathbf{1}-\mathbf{V^{'}}]-\sqrt{T}\mathbf{D}).
\end{displaymath}
If the recursion is terminated at $m=N$ $(\mathbf{A_{N+1}}=\mathbf{0})$
\begin{displaymath}
\mathbf{A_N}=\gamma^{-1}(\frac{1}{\sqrt{T}}[F\mathbf{1}-\mathbf{V^{'}}]
-\sqrt{T}\mathbf{D}),
\end{displaymath}
for $N\geq 2$. $\mathbf{C_2}$ is thus calculated as
\begin{displaymath}
\mathbf{C_2}=\frac{1}{\sqrt{2}}\mathbf{A_1A_0C_0}.
\end{displaymath}

\end{document}